# Single site-controlled inverted pyramidal InGaAs QD-nanocavity operating at the onset of the strong coupling regime

Jiahui Huang[1, †,*], Wei Liu[1,† ,*], Xiang Cheng[1], Alessio Miranda[2], Benjamin Dwir[2], Alok Rudra[2], Eli Kapon[2] and Chee Wei Wong[1,*]

[1]Mesoscopic Optics and Quantum Electronics Laboratory, University of California, Los Angeles, 420 Westwood Plaza, CA 90095, USA

[2]Institute of Physics, École Polytechnique Fédérale de Lausanne, Lausanne, VD 1015, Switzerland

*Correspondence: jiahuihuang@ucla.edu; weiliu01@lbl.gov; cheewei.wong@ucla.edu

†These authors contributed equally.

**Precise positioning of single site-controlled inverted pyramidal InGaAs QD at the antinode of a GaAs photonic crystal cavity with nanometer-scale accuracy holds unique advantages compared to self-assembled QDs and offers great promise for practical on-chip photonic quantum information processing. However, the strong coupling regime in this geometry has not yet been achieved due to the low cavity $Q$-factor based on the (111)B-oriented membrane structures. Here, we reveal the onset of phonon-mediated coherent exciton-photon interaction on our tailored single site-controlled InGaAs QD - photonic crystal cavity. Our results present a Rabi-like oscillation of luminescence intensity between excitonic and photonic components correlated with their energy splitting pronounced at small detuning. Such Rabi-like oscillation is well reproduced by modeling the coherent exchange of the exciton-photon population. The modeling further reveals an oscillatory two-time covariance at QD-cavity resonance, which indicates that the system operates at the onset of the strong coupling regime. Moreover, by using the cavity mode as a probe of the virtual state of the QD induced by phonon scattering, it reveals an increase in phonon scattering rates near the QD-cavity resonance and asymmetric phonon emission and absorption rate even around 50 K.**

**Introduction**

The strongly coupled quantum dot (QD)-cavity system has been proven as an important non-classical light generation and quantum information processing unit in an integrated quantum photonic circuit, realized by the on-demand generation of non-classical light through photon-induced tunneling and blockade via strong exciton-photon coupling [1]–[3] and spin-controlled quantum switch via strong QD spin-photon coupling [4]–[9]. In the weakly coupled case, the Purcell enhancement can improve the photon generation rate for entangled photon pairs through cavity decay channels [10]–[12]. In the coupled exciton-photon hybrid architecture, the decoherence, on one hand, fundamentally affects the information losses through the environment during the quantum information processing such as coherent control for secure quantum communications [13]–[17], quantum memory enabled single-photon switch [18]–[20], coherent emitter-emitter coupling [21]–[23], and quantum network protocol based on deterministic cluster state generation [24]–[30]. On the other hand, exciton pure dephasing [31]–[34] and phonon-induced dephasing [35]–[40] also play important roles in the QD-cavity interacting dynamics, Jaynes-Cummings nonlinearities, and intermediate coupling regimes [41].

Up to date, all the above quantum applications based on semiconductor QDs mostly rely on self-assembled QDs and suffer from random single QD locations, which hampers further optical lithography steps, scalable integration, and thus practical integrated quantum photonics [42]. However, single site-controlled InGaAs QDs grown at patterned pyramidal pits on the (111)B-oriented GaAs substrate (referred to as pyramidal QDs) hold unique advantages compared to self-assembled QDs [43] and offer remarkable promise for monolithic functional integrated quantum photonic circuits. Its deterministic QD nucleation and position control permit efficient light-matter interaction by placing multiple-QD systems at the desired antinodes of photonic crystal (PhC) cavities and waveguides with nanometer accuracy [44]. Their high structural symmetry allows reduced fine structure splitting for polarization-entangled photon generation [45],[46]. The absence of spurious cavity feeding from wetting layers [47] and parasite QDs in the neighborhood of the QD is a consequence of the peculiar MOVPE growth of QD at the sharp tip of the pyramid [48],[49]. The inhomogeneous broadening of pyramidal QD is as low as 1.4 meV thanks to precise control of the nominal pyramid size [50],[51]. This enables producing large-scale arrays of similar QD-cavity systems [46]. Furthermore, more complicated functional quantum photonic circuits requiring the incorporation of many QDs, which cannot be achieved reliably with self-assembled QDs, have been reported using site-controlled pyramidal QDs in PhC waveguides [52], tilted cavities [53], and Fano cavities [54].

Achieving the strong coupling regime and the impact of dephasing in strong exciton-photon interaction were mostly studied based on self-assembled QDs [55]–[63], but up to now it has rarely been reported on site-controlled pyramidal InGaAs QDs, which limits their usage for applications requiring strong light-matter interactions. This is because (1) previously studied pyramid InGaAs QDs exhibit s-state emission around 1.4 eV where the cavity $Q$-factor is relatively low [64]–[66]; (2) (111)B-oriented GaAs substrates and membranes, which are essential for etching highly symmetric pyramidal recesses for site-controlled QD epitaxy [44],[67], exhibit increased optical cavity losses compared to the conventional (100)-oriented substrates and membranes.

In the present study, by increasing the Indium content and reducing the pyramid nominal size, the s-state emission energy of our single site-controlled pyramidal InGaAs QD is shifted to ~ 1.24 eV without compromising its sub-100 μeV excitonic linewidth [50],[68]. At such photon energies, the reduced absorption losses from the Urbach tails of GaAs [69],[70] and smaller structural disorder thanks to larger PhC parameters lead to $Q$-factors up to ~ 12000 as compared to previous studies with lower $Q$-factor (< 4500) that were carried out in the weak coupling [64]–[66] or intermediate coupling [41] regimes. Based on our improved devices, we investigate the cavity quantum electrodynamics (cQED) at the onset of the strong coupling regime mediated by phonon induced dephasing, using cryogenic micro-photoluminescence (μPL) supported by cQED modeling. By adjusting the QD-cavity mode (CM) detuning through temperature variation, we observe their linewidth narrowing and energy splitting at resonance. A corresponding cQED model reveals the role of pure and phonon-induced dephasing in our system and confirms its operation at the onset of strong coupling. This assertion is further solidified by our experimental observation of Rabi-like oscillation and quantum beating between the upper and lower energy components by varying the pump power, which only occurs at small QD-CM detuning and can be well reproduced by our modeling of the system's initial states. Such observation of the onset of strong coupling regime mediated by phonons in our device extends the potential usage of site-controlled pyramidal InGaAs QD-PhC cavity system, previously usually operating in weak or intermediate coupling regimes. This is useful for nonlinear optics and coherent control for quantum information processing in a monolithic functional integrated quantum photonic circuit, which cannot be made reliably with self-assembled QDs.

**Optical properties and recombination dynamics**

We first characterize the µPL spectra of our site-controlled QD - L7 photonic crystal cavity (PhC) cavity system (see inset in Fig. 1(b)). Figure 1(a) presents the µPL spectra of decoupled (red) and coupled QD-cavity (black) systems measured at different sample temperatures, excited by a 900 nm pulsed diode laser with 40 µW average pump power. At 20K, a QD s-state exciton line is around 1.2445 eV with a linewidth as narrow as $92 \pm 4$ µeV. Around such photon energy, the linewidth of the fundamental $L7$ cavity mode is about $100 \pm 2$ µeV, which corresponds to $\approx 12,000$ $Q$-factor, significantly larger than previously reported values in similar (111)B-based membrane cavities [64]–[66]. We note that a low rate of excitonic dephasing and cavity dissipation, along with large coupling strength by accurately positioning the QD at the maximum of the cavity field is a prerequisite for strong exciton-photon coupling [71],[72]. The exciton emission can be resonantly coupled with cavity mode as temperature increases up to around 50 K. The Purcell-enhanced coupled QD-CM PL intensity is about 8-fold larger than the decoupled excitonic PL intensity. In the inset of Figure 1(a), by performing Hanbury-Brown-and-Twiss (HBT) measurements with a 532 nm continuous wave laser excitation, we verify the single photon emission characteristic with the degree of second-order coherence $g^{(2)}(\tau=0) \approx 0.42 < 0.5$ at QD-CM resonance, which supports that the coupled QD-CM is at the single-exciton and single-photon levels. Note that the avalanche photodiode used has a limited single photon timing resolution of 350 ps which can results in a non-zero $g(2)(\tau=0)$. Such distortion of $g(2)(\tau=0)$ can be improved using superconducting nanowire single photon detectors (SNSPDs) or resonant/quasi-resonant pumping however it is beyond our current capability. Figure 1(b) examines the polarization features of the cavity mode and excitonic emission with a detuning of $\sim 3.8$ meV (T = 35 K). Its inset presents the scanning electron microscope (SEM) image of the QD-cavity structure, where the dot-triangle sign labels the position of the inverted pyramidal QD. The luminescence through the cavity mode reveals a predominant vertical polarization resulting from the coupling of the QD emission to the TE-polarized field of the cavity mode. On the other hand, the QD excitonic emission not resonant with the cavity mode exhibits anti-polarization with respect to that of the cavity mode shown as a negative degree of linear polarization (DOLP). Such anti-polarization at large positive detuning is explained as a result of the destructive quantum interference of the free-space and cavity mode mediated decay channels through V-polarized modes [73].

We further perform the time-resolved µ-PL (TRPL) with sub-bandgap 900 nm pulsed excitation at 20 MHz repetition rate with lower average power at 10 µW in order to characterize the QD-cavity interaction dynamics at low excitation levels. Figure 1(c) manifests a shortening of the decay time from 2.5 ns of the decoupled QD excitonic decay to

1.2 ns of the resonantly coupled QD-CM PL, which indicates Purcell enhancement. The instrument response is about 0.5 ns, as verified by probing the laser pulse. In particular, the PL decay of the resonantly coupled QD-CM system reveals a slower decay at an early delay, which is likely linked to the coupled QD-CM dynamics [74]. As a comparison, the decay of decoupled excitonic PL is mono-exponential and starts at an early delay, which confirms the absence of saturation of the QD in the low injection region. Figure 1(d) further presents the coupled QD-CM PL decay (linear scale of the Y axis) as a function of pump power from 5 µW to 40 µW. Up to 20 µW, the QD-CM coupling-induced plateau at early time delay extends up to 2 ns with increasing power. This plateau cannot be attributed to multiexcitonic transitions originating from the wetting layer of quantum well (QWR) along the edges of the pyramids. The former typically observed in self-assembled QD systems is not present in the pyramidal QD systems [65],[66] and the latter reported to appear as an additional peak at early delay with fast dynamics [66] is not observed in our results. Note that the intensity of CM and QD exciton saturated around the same power suggests the absent contribution from multiexcitonic background continuum (Figure S1 in the supplementary materials). Instead, such PL plateau can be related to the QD-CM interactions. However, further measurements with finer time sampling should be performed with longer laser pulse with energy resonance with coupled QD-CM for further investigation. As the power increases to > 20 µW, a secondary raising of PL intensity appears around 2 ns, which results from the finite p-state occupation of QD by carrier feeding [66]. Such finite p-state feeding likely leads to a reduction of lifetime of coupled QD-CM decay as shown in Figure S2.

**Phonon dynamics and dephasing toward strong exciton-photon coupling regime**

We subsequently characterize the µ-PL spectra during fine-tuning of the QD-cavity coupling near resonance by varying the sample's temperature, with a much lower pumping of 3 µW to suppress free-carrier-induced charge fluctuation. QD energy changes at a faster rate following the temperature dependence of GaAs bandgap while the CM changes at a slower rate dependent on the thermal induced refractive index change of the PhC and their coupling temperature is around 47.6 K. Cavity pulling effect previously reported as a result of the interaction of phonon sideband of the self-assembled QD with the CM [75] is not visible. As shown in Figure 2(a), the phenomenological double Lorentzian fitting is performed to the PL spectrum at temperatures where the CM and QD exciton peak can be visually distinguished. However, from T = 47 K to 48.4 K, the merging of two components leads to very closed R-square values (~ 0.99) for double and single Lorentzian fits to the spectrum as shown in Figure S3 in the supplementary materials. It suggests that the double Lorentzian analysis at T = 47 to 48.4 K is not reliable although their R-square values are close to 1. As a result, the anti-crossing of QD-CM is not resolved at near zero detuning range suggests a small coupling strength in this system which will be discussed later. However, it is worth noting that the typical doublet splitting in the strong coupling regime can be concealed by pure dephasing, phonon scattering, and incoherent pumping, and thus appears as a plateau or even a single peak in our QD-cavity system [31]. Indeed, as shown in Figure 2(c), the intensity ratio of the CM and QD to their summation exhibits averaging behavior near resonance as a feature of polariton which inherit the properties from both photon and exciton [56]. Importantly, the linewidth at T = 46.7 K obtained from single Lorentzian fitting of 68.9 ± 2.5 µeV reveals clear narrowing behavior, compared to the decoupled cavity linewidth of 100 ± 2 µeV, which suggests that the QD-cavity is operating at the onset of strong coupling regime. Interestingly, the cavity and exciton linewidths as shown in Figure 2(d), closed at far-off resonance, exhibit a mutual narrowing near resonance, which indicates suppression of dissipation of both upper and lower energy components, is in contrast to the typical linewidth averaging in the strongly coupled system with generally larger cavity loss than the excitonic pure dephasing rate [56],[76].

To study the impact of the dephasing mechanism (including fluctuating environmental charges and phonon scattering) in our QD-cavity system, we perform the cQED modeling with Quantum Optics Toolbox [77] to simulate the coupled QD-cavity PL spectrum as a function of detuning. In brief, the density matrix $\hat{\rho}$ of dephasing-mediated QD-cavity interaction is described by the Lindblad master equation $\frac{d\hat{\rho}}{dt} = -\frac{i}{\hbar}[\hat{H}, \hat{\rho}] + \mathcal{L}(\hat{\rho})$, where $\hat{H} = \hbar\omega_C \hat{a}^\dagger \hat{a} + \frac{\hbar\omega_X}{2}\hat{\sigma}_z + \hbar g(\hat{a}^\dagger \hat{\sigma}_- + \hat{\sigma}_+ \hat{a})$ is the Jaynes-Cumming (JC) Hamiltonian with $\omega_C$, $\omega_X$, $\hat{a}$, and $g$ denoted as the frequency of the cavity mode, transition frequency of the QD exciton, annihilation operator of the cavity photon, and QD-cavity coupling strength, respectively. $\hat{\sigma}_{+/-/z}$ are the pseudospin operators of the two-level QD excitons. The environmental dephasing is incorporated in the Liouvillian superoperator $\mathcal{L}(\hat{\rho}) = \frac{\kappa_C}{2}\mathcal{L}_{\hat{a}}(\hat{\rho}) + \frac{\gamma_X}{2}\mathcal{L}_{\hat{\sigma}_-}(\hat{\rho}) + \frac{P_X}{2}\mathcal{L}_{\hat{\sigma}_+}(\hat{\rho}) + \frac{\gamma_{deph}}{2}(\hat{\sigma}_z \hat{\rho} \hat{\sigma}_z - \hat{\rho}) + \frac{P_{ph}}{2}\mathcal{L}_{\hat{\sigma}_- \hat{a}^\dagger}(\hat{\rho})$, where $\kappa_C$, $\gamma_X$, $P_X$, $\gamma_{deph}$, and $P_{ph}$ are the cavity loss rate, exciton decay rate, QD incoherent pumping rate, exciton pure dephasing rate causing the Lorentzian broadening of the exciton line, and phonon-mediated coupling rate from the QD excitations to the cavity, respectively [64],[78]. $\gamma_X$ is set to be 0.7 μeV which corresponds to practical spontaneous decay rate of bulk QD excitons so it's worthnoting that the phononic backscattering term describing the phonon mediated transferring of CM excitations to the QD is neglect because $\kappa_C \gg \gamma_X$ in our system [40],[64]. The superoperator $\mathcal{L}_{\hat{X}}$ with arbitrary operator $\hat{X}$ in the subscript has the form $\mathcal{L}_{\hat{X}}(\hat{\rho}) = 2\hat{X}\hat{\rho}\hat{X}^\dagger - \hat{X}^\dagger \hat{X} \hat{\rho} - \hat{\rho}\hat{X}^\dagger \hat{X}$. The experimental PL spectrum can be modeled through a steady-state power spectrum of the QD-CM system, which relates to the Fourier transform of the two-time covariance function of the excitonic operator and photon operator to be the $S(\omega) = \int_{-\infty}^{\infty} \lim_{t \to \infty} \langle A(t+\tau)B(t)\rangle e^{-i\omega\tau} d\tau$ with $A, B = \hat{\sigma}_+, \hat{\sigma}_-, \hat{a}^\dagger, \hat{a}$. Note that only a single QD is incorporated in the cavity with all surrounding QDs etched away, thus the parasitic cavity pumping in typical self-ensemble QDs is absent in our system. Figure 3 (a) shows the energies corresponding to $\omega_C$ and $\omega_X$ extracted from the modeling of the spectrum in Figure 2. In Figure 3 (b), we first adjusted $\kappa_C$ to 80 μeV deconvolved system response to reproduce spectral linewidth of the cavity at the large detuning and keep constants for all temperatures (detuning). Similarly, $\gamma_{deph}$ is set to be 30 μeV to match the QD exciton linewidth at the lower temperatures. We derive a slight increase of exciton pure dephasing $\gamma_{deph}$ at 49 K, which can be attributed to a small increase in fluctuations of the environmental charges at higher temperatures, consequently manifested in a broadening of exciton linewidth. The phonon scattering rate $P_{ph}$ is treated as a fitting parameter to reproduce the experimental spectrum shown in Figure 2 (a). The coupling strength is adjusted to be $g = 18$ μeV and kept constant for all fittings. Note that such coupling strength sets our system at the onset of the strong coupling regime ($g \approx \kappa_C/4$), which is consistent with our observation of single peak rather than explicit mode splitting. The fitted $g$ is relatively smaller than 20-50 μeV found in similar systems [64] can be attributed to a larger mode volume of the $L7$ cavity. The parameter $P_{ph}$ manifests a raising from 2 μeV when approaching the QD-CM resonance and shows a plateau with 7 μeV within ± 50 μeV detuning (grey shaded region) and subsequently falls when gradually decoupling between QD and cavity occurs, which indicates enhanced phonon scattering rate for transferring QD excitations to the cavity around near-zero detuning. Here, cavity mode can also be regarded as a sensitive probe of virtual states of QD induced by the emission or absorption of a phonon [78], which continuously spreads about 200 µeV around the QD exciton line [67]. Therefore, the elevated $P_{ph}$ within ± 50 µeV detuning is a result of convolution of the energy range of QD virtual states with the CM linewidth. It is worthnoting that the small temperature changes during the coupling (~ 2.5 K corresponding to ± 100 µeV detuning range) do not affect the phonon bath significantly which agrees with the plateau of $P_{ph}$ in such detuning range. It suggests that the pronounced $P_{ph}$ is mainly a result of QD-cavity coupling. $P_{ph}$ also exhibits an asymmetric shape between positive and negative detuning, which indicates it is more efficient for QD excitons to emit a phonon than absorb a phonon for bridging the QD-cavity energy difference. Such asymmetry suggests that the phonon bath is not fully populated around 50 K, which is different from previous theoretical predictions of site-controlled QDs operating at the weak coupling regime [64],[67]. $P_{ph}$ is also analyzed at a higher pump power of 30 µW denoted as black triangles in Figure 3(b), which shows a plateau similar to the trend at 3 µW pumping. But interestingly, $P_{ph}$ now has a larger maximum value of 12 µeV and becomes more asymmetric. Higher pump power leads to increased carrier populations and, assuming the same amount of phonon populations in our system since the coupling temperature stays the same, a higher $P_{ph}$ is required to transfer the QD excitations to the cavity decay channel. More data regarding spectral analysis at 30 µW can be found in the Supplementary Materials section VII. We expect enabling deterministically engineering QD-cavity at desired coupling temperature will open opportunities for studying temperature effects on $P_{ph}$ shape. However, challenges include degradation of QD optical quality due to larger Indium content, broadening of QD spectrum due to impurities and defects in the QD vicinity, and poor quality of cavity modes due to fabrication-induced disorder which are beyond the scope of this study.

To better illustrate the significant impact of $P_{ph}$ on phonon-assisted cavity feeding at different detuning, we present the modeled spectrum with various $P_{ph}$ at larger detuning (45.5 K) and near-zero detuning (47.6 K) in Figure 3(c) and 3(d), respectively. At 45.5 K, if not considering phonon-mediated coupling ($P_{ph} = 0$, dashed line in fig. 3(c)), the QD emission is stronger than the cavity mode, which deviates from the experiment (dot in fig. 3(c)). With slightly increasing $P_{ph}$ from 0 to 0.9 µeV, the PL intensity of QD decreases dramatically together with an increasing CM component as a result of enhanced process of phonon-mediated transferring of QD excitations to the cavity decay channel, and finally $P_{ph} = 1.8$ µeV well converges to the fitting to the experimental spectra. Similarly, at 47.6 K where QD and CM are coupled, the fitting is converged with $P_{ph} = 7$ µeV to well match both sides of spectral tails. Increasing $P_{ph}$ broaden the spectrum which is consistent with the fact that $P_{ph}$ is a decoherence process in our modeling. The high sensitivity of our modeling well captures the subtle impact of phonon scattering on the QD-CM coupling and solidifies the observed asymmetric shape. Note that the phonon effect itself does not shift the emission energy as we don't observe an energy shift in Figure 3(c) and (d). We further evaluate the steady state two-time covariance function of cavity photons $\phi(\tau) = \langle \hat{a}^\dagger(t+\tau), \hat{a}(t) \rangle$ near zero detuning as shown in the inset of Figure 3(d). It clearly shows an oscillation in $\tau$ with an exponentially damped envelope. The field correlation exhibits multiple local minimums in $\tau$ equivalent to the Rabi frequency [79],[80] and marks the coherent exchange between photon and exciton fields in our system even operating at the onset of the strong coupling regime. The dissipation and pumping due to environmental interactions manifest the exponential damping that de-correlates the exciton and photon fields. It's worth noting that if neglecting $P_{ph}$ but treating exciton pure dephasing rate $\gamma_{deph}$ as a fitting parameter to converge the fit to the experiment, the $\gamma_{deph}$ needs to be forced to increase as QD-CM is decoupling, which leads to an unphysical circumstance. The fitting to the experimental spectrum with $P_{ph} = 0$ is detailed in the Supplementary VI. These results highlight the important role of phonon-mediated coupling in our system operating at the onset of strong coupling subject to environmental interactions.

**Rabi oscillation and quantum beating of coupled spectral components**

To gain further insight into the interaction in the resonantly coupled QD-CM system, we perform pump power-dependent PL measurements with sub-bandgap 900 nm pulse pumping with fixed pulse duration at 100 ps at two different QD-CM detuning values.

Figure 4 (a) and (b) shows the PL spectrum at 5, 7, and 9 µW at a small ($\Delta_1 \approx 70$ µeV) and large ($\Delta_2 \approx 137$ µeV) detuning, respectively. As shown in Figure 5(a), by varying pump power, the PL intensity of upper and lower energy components manifests an out-of-phase oscillation up to 20 µW, which resembles the Rabi-like oscillation and indicates quantum beating between them. The three points marked by the arrow correspond to pump power in Figure 4 (a). Such oscillation vanishes beyond 20 µW likely due to decoherence from finite p state feeding. This is consistent with the power-dependent TRPL results in Figure 1(d), where significant carrier feeding above 20 µW causes the decoherence of QD-cavity interaction. The energy separation between the lower and upper energy components also exhibits an oscillatory signature correlated with the intensity oscillation.

Note that the 900 nm pulse (20MHz) used for this measurement excites carriers in the QWR which are subsequently captured by the single QD. As explained prior, in such pumping scheme of our pyramid QD-cavity system, the cavity is only feed by single QD s state at low power (< 20 µW) and is not contaminated by multiexcitonic background or QWR tails. To elucidate the oscillatory behavior, we approach the measured spectrum of our coupled QD-CM as an outcome of spontaneous decay from an initial state, which is prepared by the pulse area [79]. In brief, the initial state of the system can be set by the weighting factor $x$ in the superposition of QD exciton and cavity photon operator $x\hat{\sigma}_- + (1-x)\hat{a}$, where $x$ denotes the initial occupation percentage of QD excitons, which is complementary to the cavity photons. By varying $x$ and calculating the cavity output spectra, our model well captures the variation of the measured PL spectrum at different pump power. The three representative modeled spectra that fit the experiments at 5, 7, and 9 µW are shown as black solid curves in Figure 4(a). In Figure 5(b), the initial occupation of QD excitons is plotted as red dots with the corresponding extracted QD-cavity energy splitting $\Delta E$ as black dots from the modeling. Importantly, we note that the lower occupation of QD excitons as an initial state of the system corresponds to a large splitting between the upper and lower components, and vice versa, which leads to an out-of-phase pump power dependent oscillation of the initial occupation of exciton and $\Delta E$. Such behavior of PL intensity beating and splitting oscillation is consistent with the prior theoretical prediction of the spontaneous emission spectra [79] of the spectral shape evolution with the initial state as one exciton or one photon in a strongly coupled system. Note that complete photon domination ($x = 0$) of the system appears as pump power increases to around 4.5 µW in the first oscillation period but smears out in the following oscillation period. Such drops in the visibility at higher pump power (>20 µW) are in line with the blurred PL intensity oscillation and thus the decoherence of QD-cavity interaction due to increased carrier feeding, evidenced by the TRPL measurement in Figure 1(d). With continuing increasing power up to 480 µW, the two peaks merge and shows a single peak which suggests the system either remain in the strong coupling in disguise as a single peak [31] or enters the weak coupling regime [79],[81],[82] as shown in Figure S8. Not that such oscillation cannot be observed using a 532 nm cw pump which is shown in Figure S9 where the above GaAs bandgap pump introduces significant broadening and decoherence and the cw pump sets the system in the steady state.

The analysis of pump power-dependent PL performed at the large detuning $\Delta_2$ is shown in Figure 5(c) and (d). Clearly, the Rabi-like oscillation of the intensity and the energy separation of the upper and lower energy components, which is pronounced at smaller detuning, now blurs out. Although a weak oscillatory signature is barely observed at low pump power, the PL intensity of upper and lower components saturates at higher pump power, suggesting that the two components become photon or exciton-dominated so that QD and CM are decoupled at such detuning. This result indicates that the oscillatory behavior, only observable at small detuning, is closely related to phonon-mediated coherent exchange between exciton and photon states through vacuum Rabi splitting (VRS), which is only significant when QD and CM sufficiently interact. Regarding Figure 5(d), it is worth noting that in the large detuning case, the energy splitting between QD and CM states gradually increases with increasing pump power. As shown in Figure S7(d), such decoupling is a result of the CM red shifting.

**Conclusion**

Highly symmetric, site-controlled InGaAs QDs formed in patterned inverted pyramid on GaAs (111)B-oriented substrates combined with deterministic PhC cavity integration offers great promise for integrated quantum photonics. However, this geometry is subject to increased cavity loss in contrast to equivalent systems involving (100)-oriented membranes and self-assembled QDs. In the present study, by tailoring the Indium content and pyramid nominal size, we demonstrate and investigate cQED in single site-controlled pyramidal InGaAs QD-PhC cavity systems operating at the onset of the strong coupling regime. Further cQED modeling of the power spectra accounting for environmental dephasing reveals the pronounced phonon-mediated-coupling rate and an oscillatory two-time covariance function near zero QD-cavity detuning. This indicates that our phonon-mediated QD-cavity system, even at the onset of strong coupling, supports the coherent exchange of exciton and photons. It is further substantiated in the pump pulse area-dependent PL measurement that Rabi-like out-of-phase oscillation and quantum beating between emission intensity of the upper and lower energy components occurs, which is also correlated with an oscillatory feature of their energy splitting, taking place only at small QD-cavity detuning and lower pump power. The oscillation of energy splitting of upper and lower components can be well reproduced by modeling the initial occupation of excitons in the system. Our results demonstrate the feasibility of using single site-controlled pyramidal InGaAs QD near the strong coupling regime and provide guidelines for further optimizing the fabrication process to further reduce the cavity loss in this geometry. It enables and extends the applications of such systems to schemes for coherent control of site-controlled exciton and photon states for quantum information processing in monolithic functional integrated quantum photonic circuitry for practical applications.

**Supplementary material**

The fabrication of site-controlled inverted pyramidal InGaAs QD – *L*7 cavity, µPL measurement schematic, and additional experimental data analysis can be found in the accompanying supplementary material.

**Acknowledgment**

The authors thank helpful discussions with Murat Can Sarihan, Abhinav Kumar Vinod, James F. McMillan, Yujin Cho, Kai-Chi Chang, Jin Ho Kang, Justin Caram, and Baolai Liang from UCLA and technical help from Alexey Lyasota and Bruno Rigal from EPFL in the sample fabrication. The authors also acknowledge exchange discussions with Sophia Economou and Bikun Li. J.H., W.L., and C.W.W. acknowledge support from the Army Research Office MURI (W91INF-21-2-0214) and National Science Foundation (2137984, 1936375, and 1919355). W.L. also acknowledges support from the Swiss National Science Foundation under project 187963. W.L. and J.H. led the project and performed the measurements with data analysis and cQED modeling. X.C supported the cQED modeling. B.D. and E.K. designed the samples. A.M., B. D., and A. R. grew the samples and performed the photonic crystal microcavity fabrication. C.W.W. and E.K. aided in the project. W.L., J.H., and C.W.W. wrote the manuscript, with contributions from all authors.

**Conflict of interest**

The authors have no conflicts to disclose.

**Data availability**

The data that support the findings of this study are available from the corresponding authors upon reasonable request.

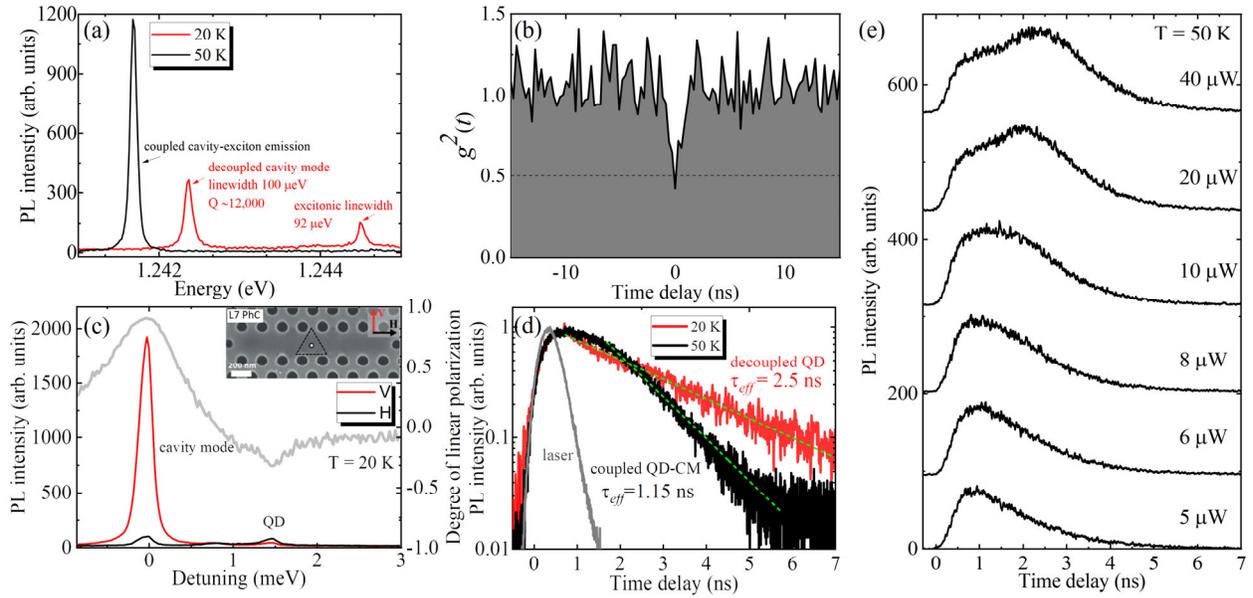

**Figure 1 Optical properties and recombination dynamics (a)** μ-PL spectra of decoupled (red) and coupled QD-cavity (black) system. **(b)** Second-order correlation function of the coupled QD-cavity emission. **(c)** μ-PL of TE and TM cavity mode and QD single excitonic line measured at T = 20 K, with vertical (red) and horizontal (black) polarization. The grey line indicates the degree of linear polarization $P = (I_V - I_H)/(I_V + I_H)$. Inset: SEM image of single site-controlled QD - L7 photonic crystal cavity, where the position of pyramidal QD, as dot-triangle labeling, matches the antinode of the cavity field distribution. **(d)** PL decay of decoupled excitonic emission (red) at T = 20 K and coupled QD-CM (black) at T = 50 K, corresponding to 2.46 ns and 1.15 ns decay time, extracted by single exponential decay, respectively. The grey decay curve records the laser decay, which suggests an about 500 ps instrument response. Measurement is performed with an average power of 40 μW at 80MHz repetition rate with a 900 nm pulse laser. **(e)** Time-resolved PL traces of the resonantly coupled QD-CM as a function of pump power at T = 50 K. Measurement is performed at 20MHz repetition rate with a 900 nm pulse laser.

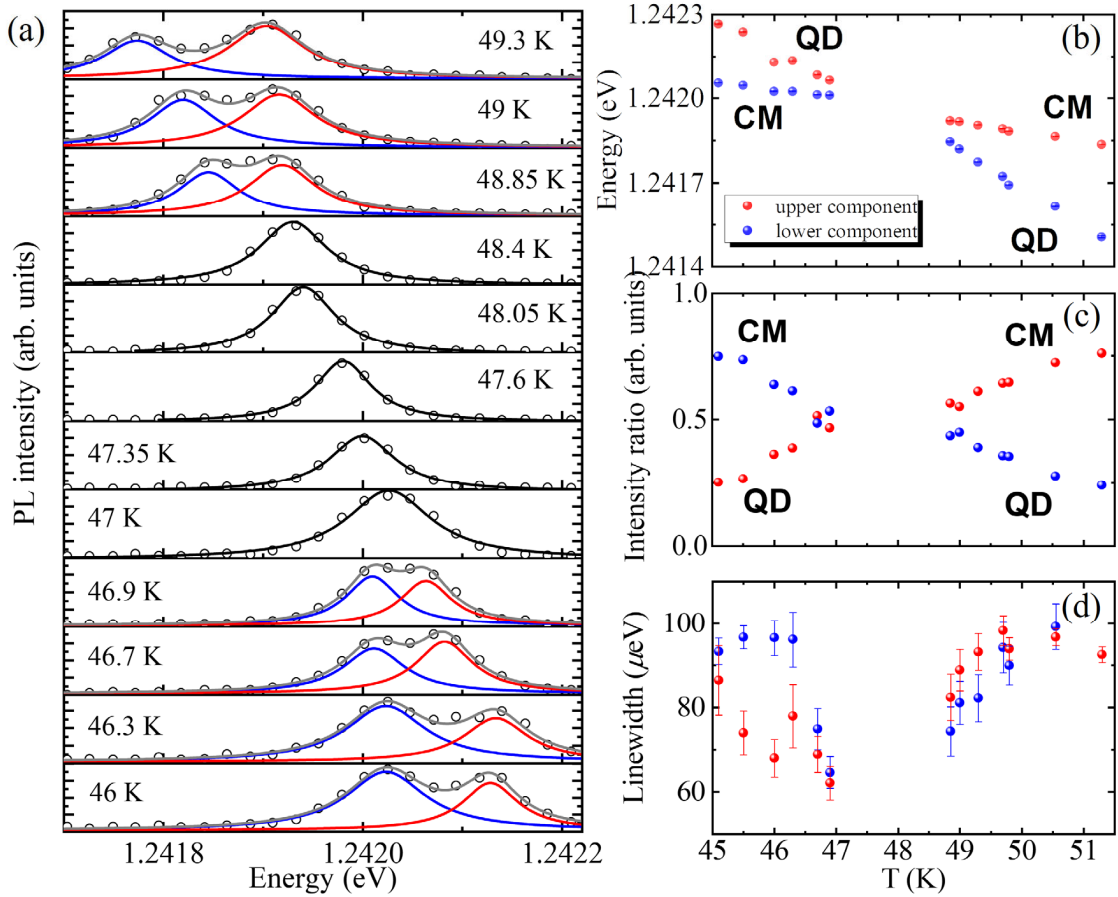

**Figure 2 Temperature tuning of QD-cavity coupling (a)** Experimental PL spectra (circle) of the coupled QD and CM around resonance fine-tuned by varying temperature. The red and blue curves represent the double Lorentzian fitting to the upper and lower energy components, respectively. The grey curve is their cumulative fit. The black curves at 47 K to 48.4 K are the single Lorentzian fitting. **(b)** Dependences of PL peak energies of coupled QD-cavity system over a wide temperature range. **(c)** Corresponding detuning dependent PL intensity ratio between the upper and lower energy components to their summation, which reveals averaging near resonance. **(d)** Corresponding detuning-dependent linewidth of the upper and lower energy components.

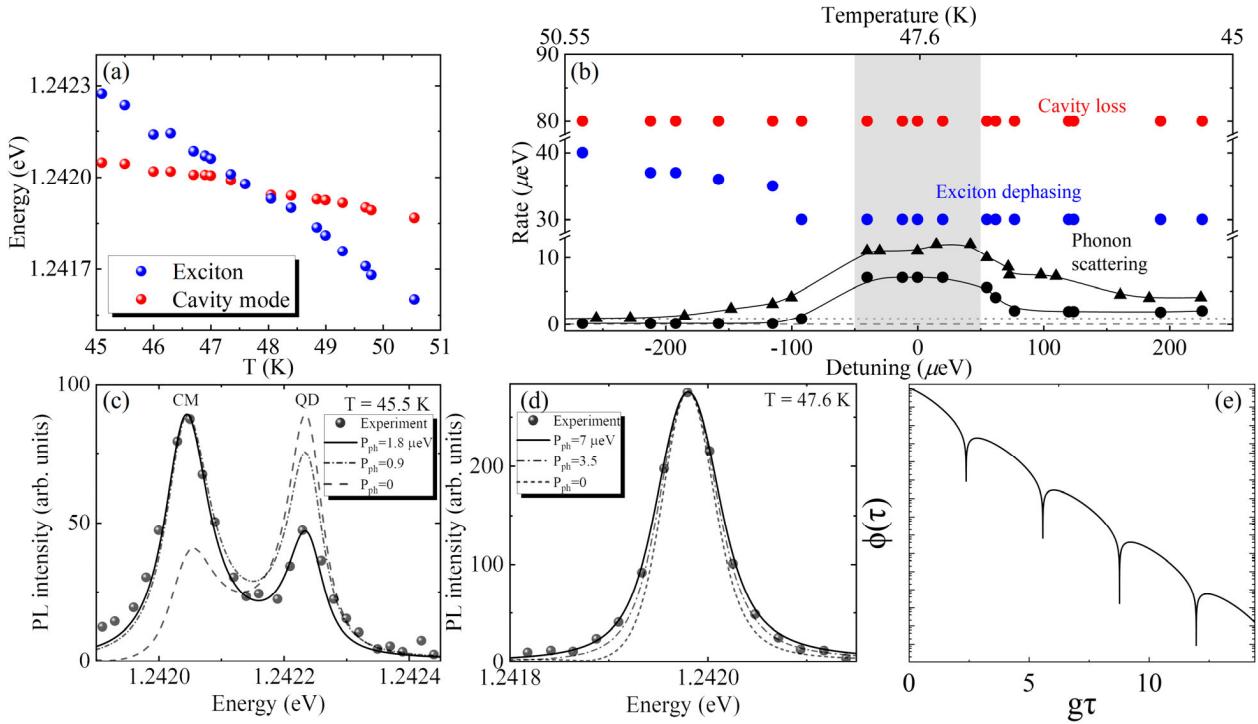

**Figure 3 cQED modeling of phonon-mediated QD-cavity coupling (a)** Fitting of energy of exciton (blue) and cavity mode (red) as a function of temperature in Figure. 2 (a) by modeling. **(b)** Cavity loss (red), exciton dephasing (blue), and phonon scattering (black) rate as a function of temperature in the QD-cavity system extracted from the modeling in **(a)**. Black triangles denote the phonon scattering rate at a higher pump power at 30 µW. The grey dashed and dotted lines mark the phonon scattering rate at the large negative detuning, indicating an overall asymmetric shape across ± 200 µeV detuning range. **(c)** PL spectra (dot) of the decoupled QD-cavity system at 45.5 K. The dash, dot-dash, and solid lines are the fitting corresponding to the phonon mediated QD-cavity coupling rate $P_{ph}$=0, 0.9, and 1.8 µeV, respectively. **(d)** PL spectra (dot) of the resonantly coupled QD-cavity system at 47.6 K. The dash, dot-dash, and solid lines are the fitting corresponding to the phonon mediated QD-cavity coupling rate $P_{ph}$=0, 3.5, and 7 µeV, respectively. **(e)** Time evolution of the two-time correlation for spectra in (d).

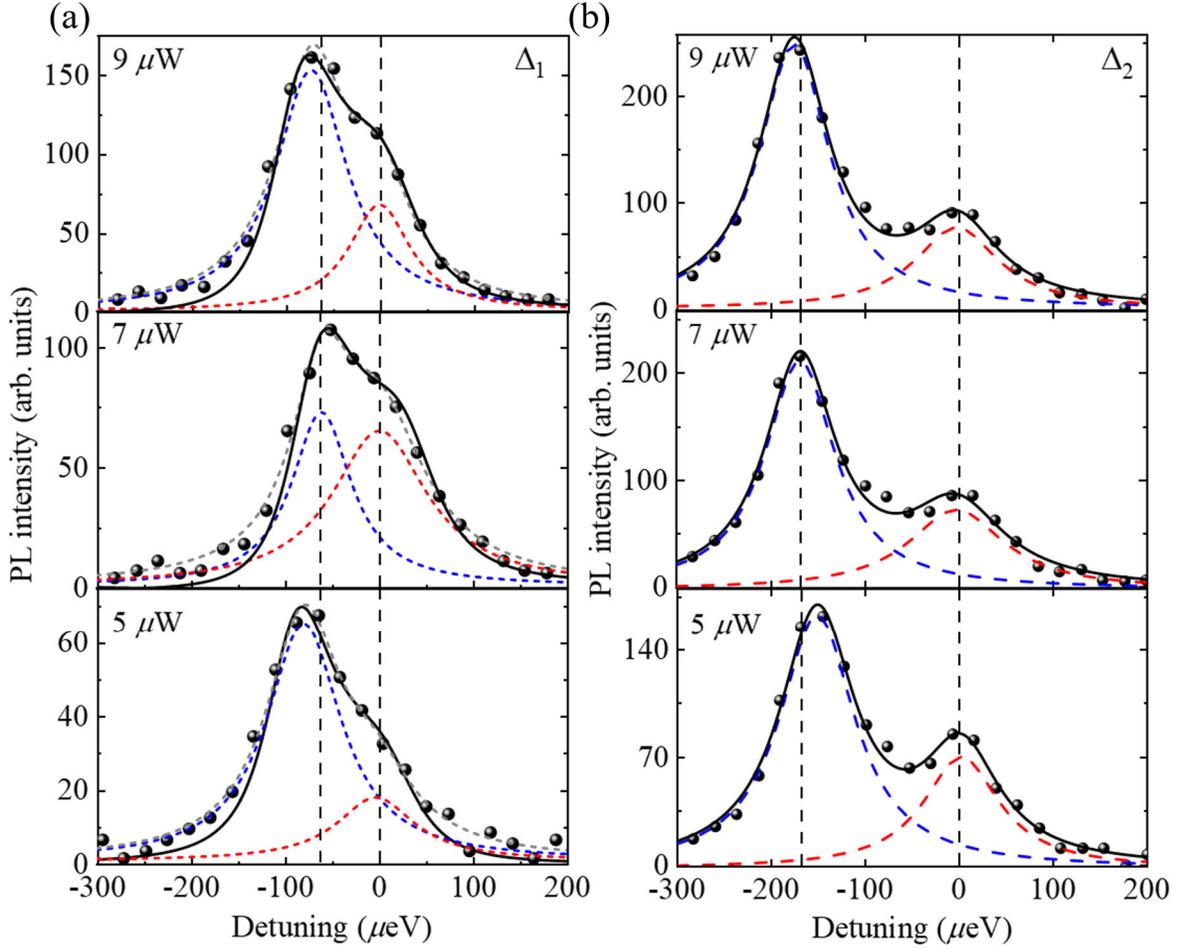

**Figure 4** PL spectra (black sphere) of the nearly resonantly coupled QD and CM excited by the 900nm pulsed laser at average power 5 μW, 7 μW, and 9 μW for **(a)** smaller ($\Delta_1 \approx 70$ μeV) and **(b)** larger ($\Delta_2 \approx 137$ μeV) QD-CM detuning cases at T = 50 K and 52.3 K, respectively. The blue and red dash lines are the double Lorentzian fitting with the accumulative fitting as the grey dashed line. The black solid line in **(a)** is from the cQED modeling. The vertical dashed line marks the peak center energies of lower and higher energy components at 7 μW.

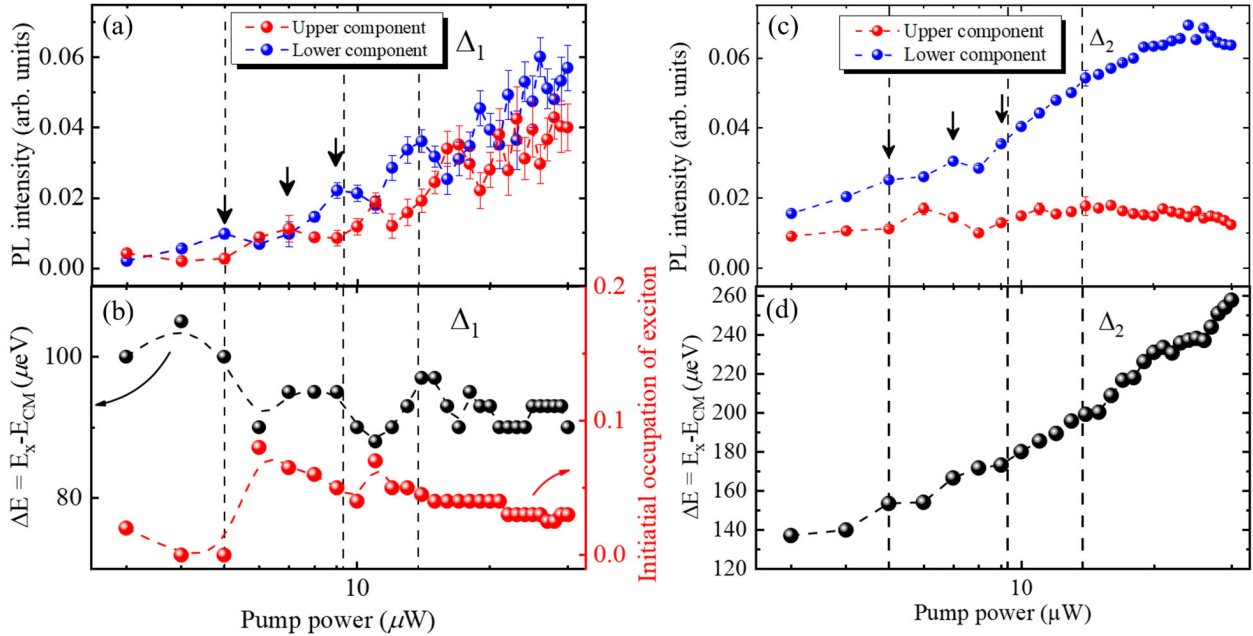

**Figure 5 Power dependent Rabi-like oscillation of the coupled QD-cavity (a)** PL intensity of upper and lower energy components as a function of average pump power extracted for the small detuning ($\Delta_1 \approx 70$ µeV) at T = 50 K, which reveals clear oscillation under low pump power. The three arrows mark the pump power at 5 µW, 7 µW, and 9 µW. **(b)** Energy separation between upper and lower energy components (black) and initial occupation of exciton in the coupled QD-cavity system (red) for the small detuning case as a function of pump power extracted from the fitting by cQED modeling. The black and red dash curves are the guide to the eye. The vertical dashed lines are guides to the eye marking the oscillation period. **(c)** PL intensity and **(d)** energy separation as a function of average pump power for a larger detuning ($\Delta_2 \approx 137$ µeV) at T = 52.3 K. The vertical dashed lines are copied from (a) and (b).

# Supplementary Materials
# Single site-controlled inverted pyramidal InGaAs QD-nanocavity operating at the onset of the strong coupling regime

Jiahui Huang[1,†], Wei Liu[1,†,*], Xiang Cheng[1], Alessio Miranda[2], Benjamin Dwir[2], Alok Rudra[2], Eli Kapon[2] and Chee Wei Wong[1,*]

[1]Mesoscopic Optics and Quantum Electronics Laboratory, University of California, Los Angeles, 420 Westwood Plaza, CA 90095, USA

[2]Institute of Physics, École Polytechnique Fédérale de Lausanne, Lausanne, VD 1015, Switzerland

*Correspondence: weiliu01@lbl.gov; cheewei.wong@ucla.edu

†These authors contributed equally.

**I. Sample fabrication and measurement schematic**

The single site-controlled pyramidal QDs are fabricated by metalorganic vapor-phase epitaxy (MOVPE) growth of $In_xGa_{1-x}As$/GaAs (x = 0.25) on the (111)B-oriented GaAs substrate with electron-beam lithography (EBL) (precision within ± 5 nm) inverted pyramidal pits pattern with a triangular lattice of pitch = 450 nm. Without the formation of 2D wetting layers, the lens shape QD is grown at the apex of a highly symmetric inverted pyramid with well-defined (111)A Gallium terminated facets (nominal pyramid size ≈ 180 nm). InGaAs/GaAs quantum wires (QWR) can be formed on the three wedges of the inverted pyramid during growth in some circumstances. *L*7 photonic crystal (PhC) structures with slab thickness t ≈ 265 nm, lattice constant a ≈ 225 nm, and air hole radius r ranging from 30 nm to 50 nm are lithographically written on top of the QD pattern and all the surrounding QDs are etched away except the one at the center of the cavity. The first three side holes on the left and right of the cavity are shifted outwards by 0.23a, 0.15a, and 0.048a. The details of QD growth and PhC cavity integration can be found in ref [S1] and [S2].

For the cryogenic micro-photoluminescence (μPL) measurement, the sample is mounted on the cold finger of a liquid helium flow cryostat, positioned accurately with piezoelectric actuators in the XY-direction. The sample is excited by a 900 nm pulsed diode laser with ≈ 100 ps pulse duration and 2 MHz repetition rate. A 100× microscopy objective with a numerical aperture (NA) of 0.7 allows a ≈ 1 μm diffraction-limited pump beam spot size on the sample. The μPL signals are collected and collimated by the same objective and refocused onto the entrance slit of a 1-meter spectrometer with 1200 grooves/mm grating, enabling a high spectral resolution of 30 pm (40 μeV) and a liquid nitrogen-cooled charge-coupled device (CCD) is used for the PL spectrum acquisition. After spectral filtering by tunable short and long-pass filters, the signal can also be fed into a single-mode fiber-based Hanbury-Brown and Twiss (HBT) setup comprising a

50:50 fiber-based beamsplitter and two Si single-photon counting modules (SPCMs) with 25-Hz dark counts. Time-correlated single-photon counting (TCSPC) is implemented for acquiring the second-order correlation (g2) and time-resolved PL (TRPL) measurements. Polarization-resolved µPL is performed by implementing a half-wave plate and a linear polarizer in front of the spectrometer entrance. The linear polarizer is fixed at the polarization orientation which corresponds to the maximum reflectivity of the grating spectrometer as well as the cavity main polarization direction.

**II. Power dependent PL intensity of decoupled QD exciton and CM**

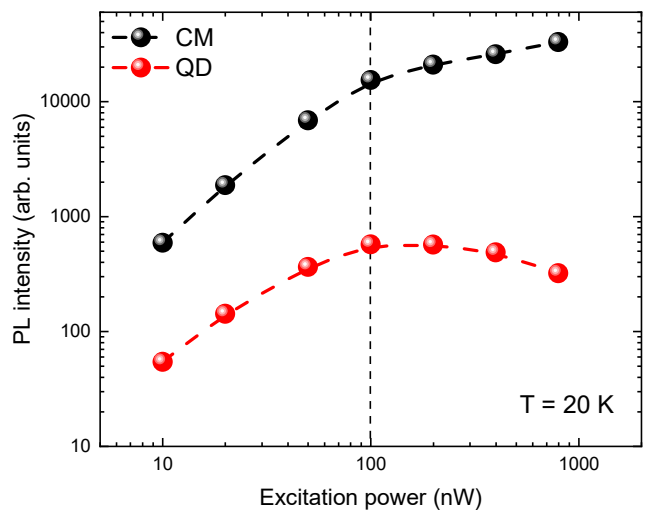

**Figure S1** Power dependent PL intensity of decoupled QD exciton and CM at T=20K using a 532 nm cw laser.

**III. Power dependent lifetime of coupled QD-cavity**

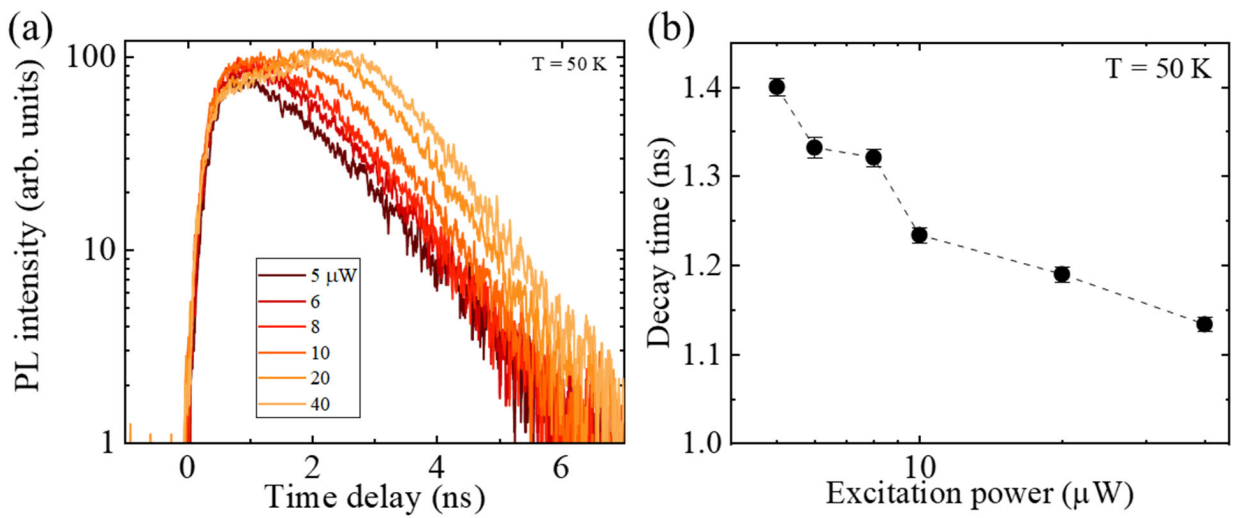

**Figure S2 power dependence of decay time (a)** PL decay of coupled QD-CM (black) at T = 50 K with pump power from 5 µW to 40 µW with PL intensity axis in log scale. Measurement is performed at 20MHz laser repetition rate. **(b)** Extracted lifetime as a function of excitation power.

## IV. Comparison between single and double Lorentzian fitting of the PL spectrum measured at 47 K – 48.05 K

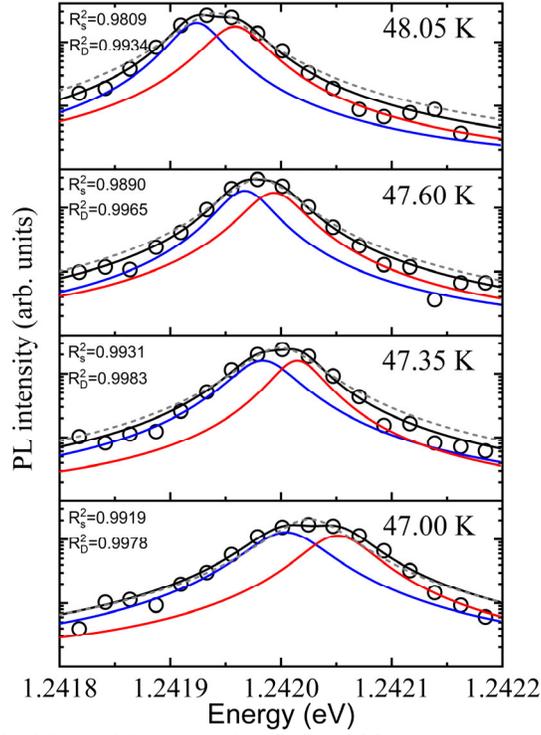

**Figure S3** Single (grey dash) and double (solid) Lorentzian fitting of four temperatures marked by hollow black circles in Fig 1 (b). R-square values for single and double Lorentzian fitting are denoted by $R_S^2$ and $R_D^2$, respectively.

Figure S1 shows the single (grey dash line) and double (solid line) Lorentzian fitting of the PL spectrum measured at 47 K to 48.05 K from figure 2(a). Both fits based on the Levenberg Marquardt algorithm converge to the experimental data without fixed fitting parameters. It can be clearly seen that the double Lorentzian line shape fits better at the tail of the experimental data and mimics the observed spectra shape.

## V. cQED modeling of coupled QD-cavity PL spectra at 48.05 K

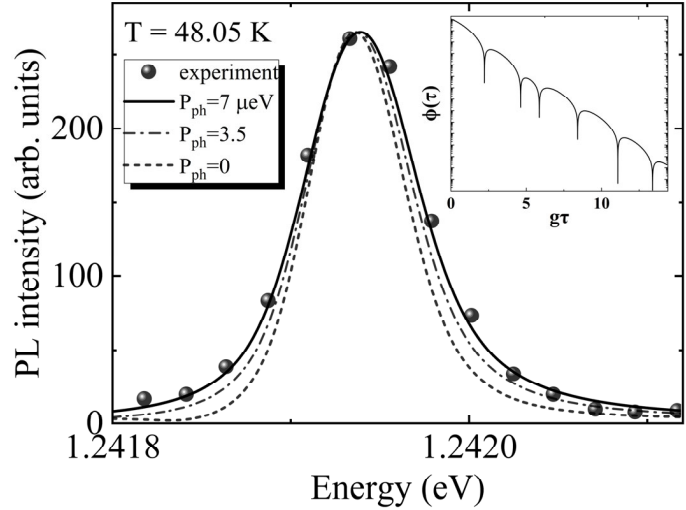

**Figure S4** PL spectra (dot) of the resonantly coupled QD-cavity system at 48.05 K. The dash, dot-dash, and solid lines are the fitting corresponding to the phonon-mediated QD-cavity coupling rate $P_{ph}$=0, 3.5, and 7 µeV, respectively. Inset: The corresponding time evolution of the two-time covariance function.

Figure S4 shows the PL spectra (dot) of the resonantly coupled QD-cavity system at 48.05 K. An increased phonon scattering rate of $P_{ph}$ = 7 µeV is required to reproduce the experimental PL spectra, similar to T = 47.6 K in Figure 3(d). It's interesting to note that the two-time covariance function corresponding to 48.05 K shows more oscillations than 47.6 K while decaying which suggests that 48.05 K corresponds to a stronger exciton photon coupling.

## VI. cQED modeling of QD-cavity coupling without the phonon scattering

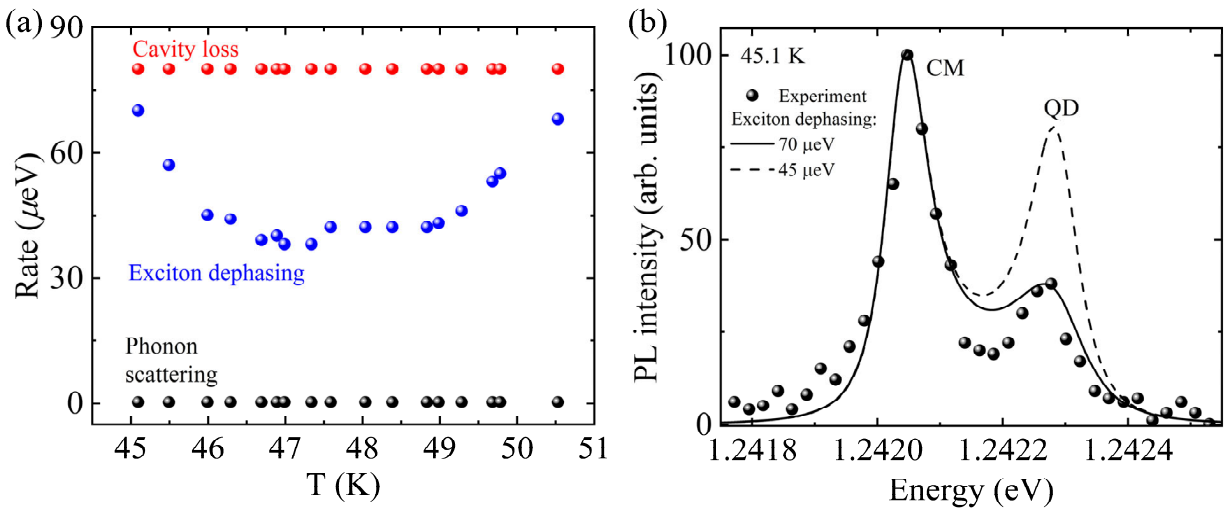

**Figure S5** cQED modeling of QD-cavity coupling without the phonon scattering **(a)** Cavity loss and exciton pure dephasing rate as a function of temperature with a vanishing phonon scattering rate. **(b)** Fitting of the measured PL spectra at 45.1 K (dot) by the cQED model with exciton pure dephasing at 70 µeV (solid line) and 45 µeV (dash line) in the case that phonon scattering rate is set to be 0.

Figure S5 shows the cQED modeling of the same QD-cavity system modeled in Figure 3 without phonon-mediated coupling. The cavity loss $\kappa_c$ is taken to be 80 µeV which is the same as in Figure 3 but now the phonon scattering rate $P_{ph}$ is set to 0 µeV for all temperatures. For the PL spectrum near zero detuning, the exciton pure dephasing rate $\gamma_{deph}$ must be reduced to converge the fitting. An example of the modeling at T = 45.1 K, where QD and CM are out of resonance, is shown in Figure S5(b) where the curve with a larger $\gamma_{deph}$ = 70 µeV (solid) reproduces the experimental spectra better than $\gamma_{deph}$ = 45 µeV (dash) but still cannot completely reproduce the experimental spectra of QD. It's worth noting that increased $\gamma_{deph}$ reduces QD exciton excitations but keeps the CM excitations constant and does not correspond to a physical circumstance. It suggests the necessity of incorporating a non-zero phonon scattering rate in the modeling which is important in the QD-cavity coupling.

## VII. cQED modeling of QD-cavity coupling at 30 µW

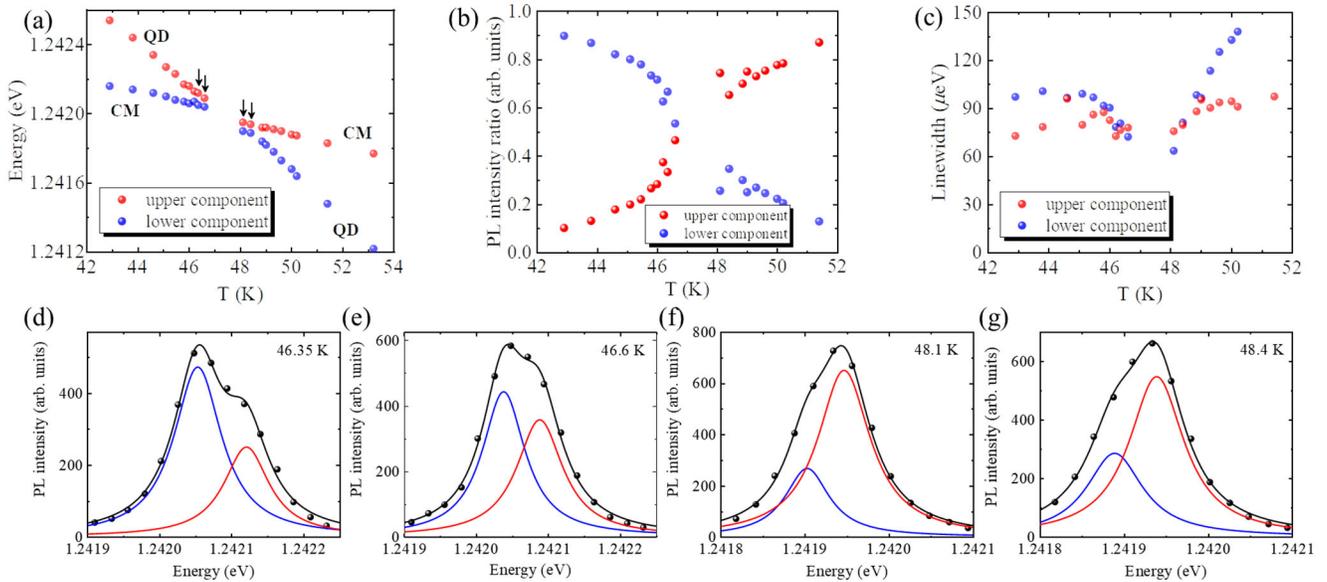

**Figure S6-1 Temperature tuning of QD-cavity coupling at 30 µW. (a)** Dependences of PL peak energies of coupled QD-cavity system over a wide temperature range. **(b)** Detuning dependent ratio of PL intensity between upper and lower energy components to their summation, which reveals averaging at near resonance. **(c)** Detuning dependent linewidth of the upper and lower energy components, which presents mutual narrowing. **(d-f)** Double Lorentzian fitting of PL spectrum at four representative temperatures marked by the arrows in **(a)**.

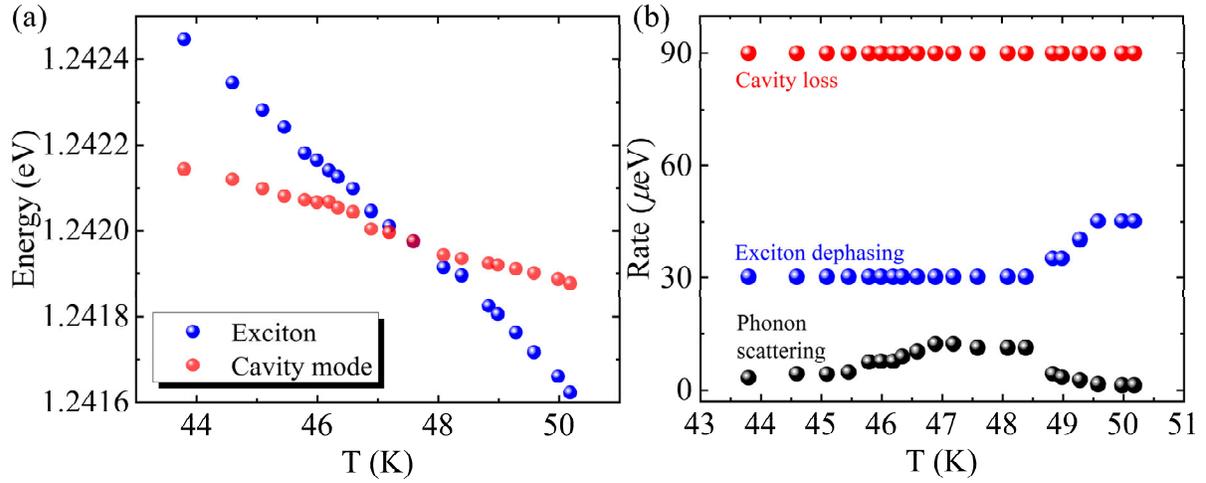

**Figure S6-2 cQED modeling of phonon-mediated QD-cavity coupling at 30 µW (a)** PL peak energies of coupled QD-cavity system as a function of temperatures extracted from the cQED modeling. **(b)** cavity loss, exciton pure dephasing, and phonon scattering rate as a function of temperatures extracted from the cQED modeling.

The temperature-dependent QD-cavity PL is also measured at a higher pump power at 30 µW. As shown in Figure S6-1 (a-c) the double Lorentzian fitting to the spectrum shows a PL intensity ratio averaging and mutual linewidth narrowing of the upper and lower components which is similar to the lower pump power case in Figure 3. The double Lorentzian fitting near zero detuning is shown in Figure S6-1(d-f). To demonstrate the robustness of our modeling and further investigate the phonon scattering rate, we fit the PL spectrum at 30 µW. In Figure S4-2(b), the detuning-dependent phonon scattering rate is asymmetric and shows a plateau from 46.9 K to 48.4 K closed to zero detuning similar to Figure 3(b). But interestingly, it has a larger maximum value of $P_{ph}$ = 12 µeV than the lower pump power case of 7 µeV. It can be due to the reason that higher pump power leads to increased carrier populations and assuming the same amount of phonon populations in our system since the coupling temperature stays the same, a higher phonon-mediated coupling rate is required to transfer the QD excitations to the cavity decay channel. Similar to the low pump power case, the exciton dephasing rate starts to increase around T = 49 K.

## VIII. Powe-dependent µPL of the coupled QD-cavity

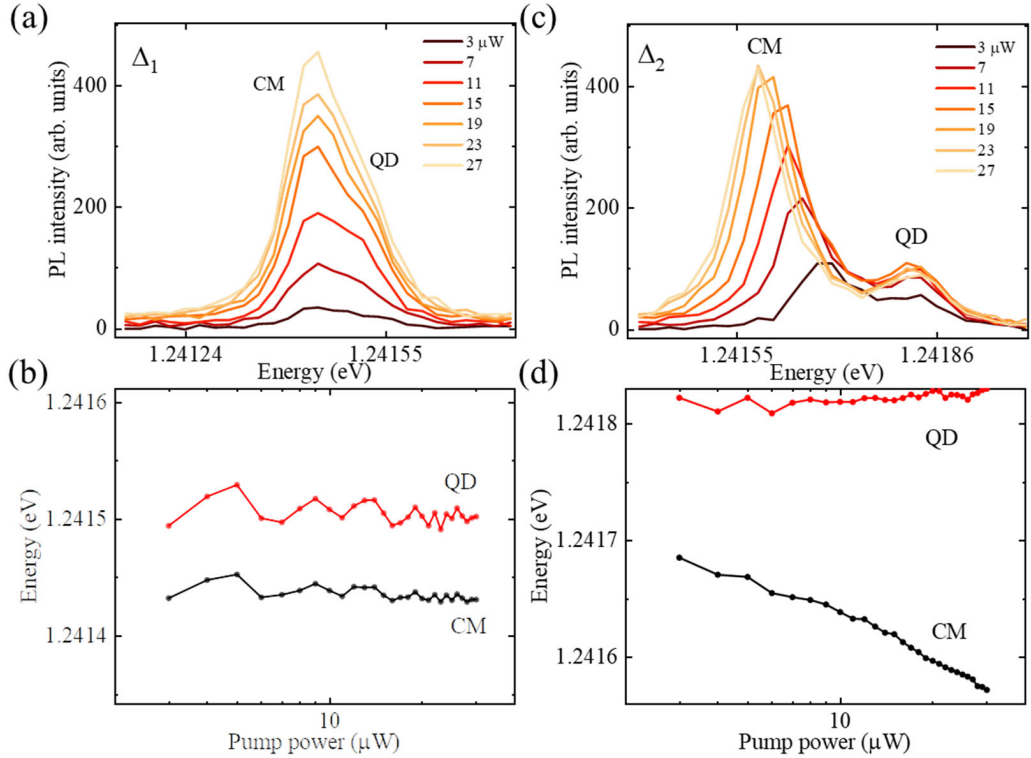

**Figure S7 (a)** Power dependent PL spectra of QD-CM for $\Delta_1 \approx 70\mu W$. **(b)** Emission energy of QD and CM as a function of pump power extracted from (a). **(c)** Power dependent PL spectra of QD-CM for $\Delta_2 \approx 137\mu W$. **(d)** Emission energy of QD and CM as a function of pump power extracted from (c).

The power dependent red shift of CM in Figure S7(c)(d) is unlikely due to the thermally induced refractive index change of the PhC due to local laser heating because (1) the 900 nm pumping at low power only excites finite carriers in the 3 QWR along the wedges of the pyramid (2) QD energy which shifts with temperature at a faster rate than CM remain constant. (3) no CM red shift is observed at 70 µeV detuning at the same power range.

PhC surface condensation of impurities due to Xenon [3] or nanofluid can lead to CM shift [4],[5]. However, none of them applied to our case because the measurement is performed when the cryostat is at thermal equilibrium so that any unintentional impurities already condense on the sample surface. Importantly, the measurements at two detuning were conducted under the same conditions. The cause of the observed red shift is currently unclear and needs further investigation.

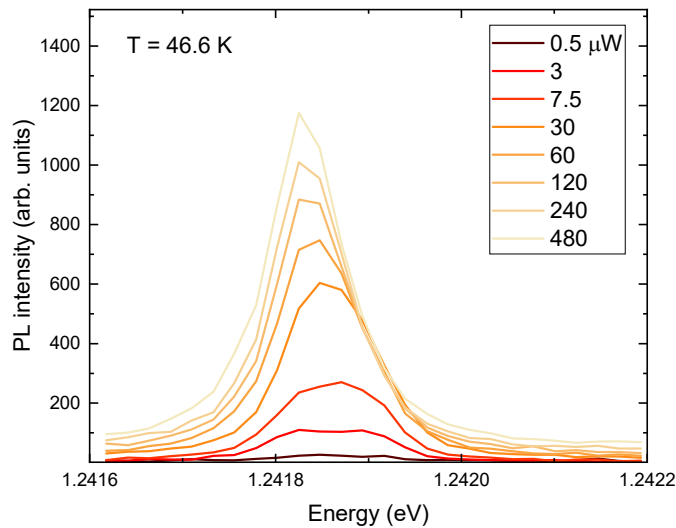

**Figure S8** PL spectrum at T = 46.6 K as a function of pump power with 900 nm laser (20MHz) up to 480 µW.

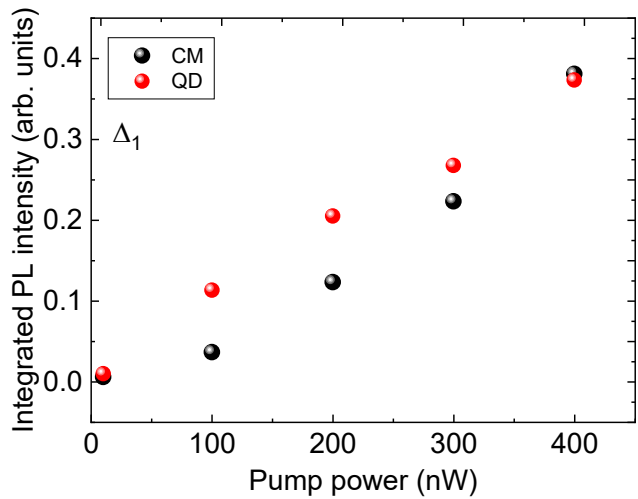

**Figure S9** Integrated PL intensity of QD and CM as a function of pump power for the small detuning case ($\Delta_1 \approx 70 \mu eV$) using 532 nm cw laser.